# A continuous contact force model for impact analysis in multibody dynamics

Jie Zhang[a], Wenhao Li[b, c], Lei Zhao[a], Guangping He[a, *]

a. School of Mechanical and Materials Engineering, North China University of Technology, Beijing, China
b. Institute of Mechanics, Chinese Academy of Sciences, Beijing, China
c. School of Engineering Science, University of Chinese Academy of Sciences, Beijing, China
*Corresponding author Email: hegp55@126.com

**Abstract**
A new continuous contact force model for contacting problems with regular or irregular contacting surfaces and energy dissipations in multibody systems is presented and discussed in this work. The model is developed according to Hertz law and a hysteresis damping force is introduced for modeling the energy dissipation during the contact process. As it is almost impossible to obtain an analytical solution based on the system dynamic equation, an approximate dynamic equation for the collision system is proposed, achieving a good approximation of the system dynamic equation. An approximate function between deformation velocity and deformation is founded on the approximate dynamic equation, then it is utilized to calculate the energy loss due to the damping force. The model is established through modifying the original formula of the hysteresis damping parameter derived by combining the energy balance and the law of conservation of linear momentum. Numerical results of five different continuous contact models reveal the capability of our new model as well as the effect of the geometry of the contacting surfaces on the dynamic system response.

Nomenclature

| | |
|---|---|
| $i, j$ | impactor, target |
| $v_i^{(-)}, v_j^{(-)}$ | initial velocity |
| $v_i^{(+)}, v_j^{(+)}$ | separation velocity |
| $t^{(-)}, t^{(+)}, t_m$ | time of initial contact, time of separation, time of maximum indentation |
| $\delta, \delta_m$ | deformation or indentation, maximum indentation |
| $\dot{\delta}, \ddot{\delta}$ | indentation velocity, indentation acceleration |
| $m$ | equivalent mass |
| $F_N$ | normal contact force |
| $k$ | equivalent stiffness |
| $n$ | non-linear power exponent |
| $c_r$ | coefficient of restitution |
| $\lambda$ | hysteresis damping factor |
| $\dot{\delta}^{(-)}, \dot{\delta}^{(+)}$ | initial indentation velocity, relative separation velocity |
| $\hat{\dot{\delta}}, \hat{\dot{\delta}}_c, \hat{\dot{\delta}}_r$ | equivalent velocity, equivalent velocity of compression phase, equivalent velocity of restitution phase |
| $\Delta E_{loss}$ | energy loss |
| $\Delta E_c, \Delta E_r$ | energy loss for compression or restitution phase calculating based on the system |

| | dynamic equation |
|---|---|
| $\Delta E_c^*, \Delta E_r^*$ | energy loss for compression or restitution phase calculating based on the approximate dynamic equation |

## 1. Introduction

Contact-impact phenomena frequently occur in multibody systems, mainly due to the clearances in bodies and joints [1-4]. Proper modeling of the contact-impact phenomenon is very important for an accurate description of the dynamic behavior of multibody systems [5]. In the past few years, the research of impact analysis in multibody systems has increased significantly [1, 6, 7]. The contact-impact process is featured by extremely short duration, large contact force, fast energy dissipation, and great changes in the velocities of bodies [8]. The modeling of contact–impact problems rely heavily on several factors, such as the topological properties of contacting surfaces, material characteristics, initial velocities and friction [9]. In order to describe the dynamic behavior of multibody system, several crucial aspects of the modeling of impact are needed to be taken into account, including the velocity change before and after the impact, the peak contact force, the duration time and the indentation depth [6].

The earliest model of impact is the coefficient of restitution, which can describe the changes in velocity and energy before and after the impact. There are different definitions of the coefficient of restitution [10, 11]. Rooted in the speed before and after the impact, Newton's definition is the most popular and commonly used among them [7]. Restitution coefficient, which is easy to measure, gives rise to a concise description of impact phenomenon. But the details of contact force and deformation in the process of collision can not be described by the coefficient of restitution, except for the velocity change and the energy loss before and after the impact.

The second approach is the nonsmooth method, in which the duration of the collision is ignored and the impact is assumed to be occurred instantaneously [12]. There are two ways to treat the contact–impact problems in a multibody system, namely the linear complementarity problem (LCP) [13, 14] and the differential variational inequality (DVI) [15, 16]. Compared with the coefficient of restitution, the nonsmooth approach can calculate the contact force with relatively efficient calculation; however, it is not valid for the modeling of impact duration due to the instantaneous assumption [9, 17].

The third approach is named as compliant continuous contact force model, owing to the contact force is described as a continuous function of the indentation depth (relative deformation). Time-varying values of the velocities, contact forces, deformations and the duration time can be depicted by this method. The nonlinear Hertz contact model proposed in 1880 has provided an important basis for fundamental research of contact mechanics until now [18]. In recent years, several continuous contact models considering the energy dissipation of impact duration have been proposed rooted in the Hertz model [19, 20], like the influential models raised by Hunt and Crossley [21] and by Lankarani and Nikravesh [17] and by Flores et al. [8]. Other continuous contact models are also developed, such as the Herbert and McWhannell model [22], the Lee and Wang model [23], the Gonthier et al. model [24], the Zhiying and Qishao model, [25] and Hu and Guo Model [9]. In these models, the hysteresis damping factor is derived as a function of restitution coefficient which is easy to measure. The strengths and weaknesses of these models have been discussed in literatures [1, 6], especially the application limitations.

A new continuous contact force model, which is inspired by the work of Flores et al. [7] and Hu and Guo [9], is proposed in this paper. The new model is expected to solve contact problems with regular or irregular contacting surfaces, which contains the contact events between soft materials with low or medium values of restitution coefficient. The remainder of the paper is organized as follows.

Section 2 covers the basic knowledge of the continuous contact force models. Then the energy loss associated with the restitution coefficient is described in Section 3. Section 4 demonstrates the construction of the new contact model. Numerical simulations and conclusions are presented in Sections 5 and 6 respectively.

**2. General issues of continuous contact force models**

As shown in Fig.1a, two solid objects (with masses $m_i$ and $m_j$) having a direct-central impact and the deformation taking place in the local contact zone, the whole process is divided into two phases: the compression phase and the restitution phase. At the initial moment of impact $t^{(-)}$, the objects have velocities $v_i^{(-)}$ and $v_j^{(-)}$, then the deformation increases until the relative normal deformation between the contacting bodies reaches the maximum $\delta_m$ at time $t_m$, after that the deformation begins to recover, the contact force gradually decreases until it becomes zero at time $t^{(+)}$, and the velocities become $v_i^{(+)}$ and $v_j^{(+)}$. The period from $t^{(-)}$ to $t_m$ is the compression phase and the restitution phase starts from $t_m$ and ends at $t^{(+)}$, as shown in Fig.1a. When the size of the contact area is much smaller than the sizes of the contacting objects, the contact-impact system with two objects can be equivalent to a collision between an object with equivalent mass $m$ and an elastic half space [7, 26], as illustrated in Fig.1b, this equivalence is valid for most elastic contact-impact problems [26, 27].

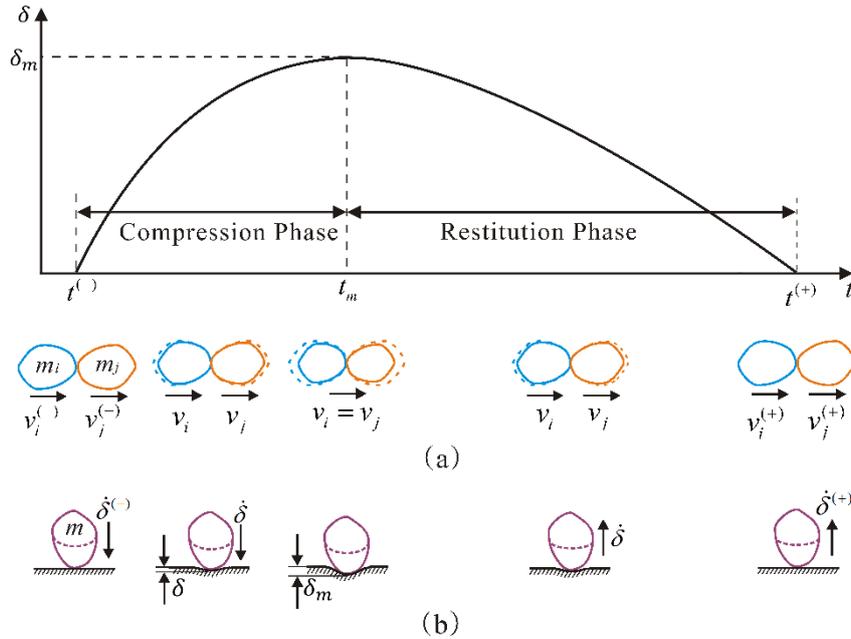

Fig.1 Contact process between two impacting objects and the equivalent system.

The groundbreaking work of Hertz on contact problems has remained an important basis for both fundamental research and engineering application of contact mechanics until now. Based on the elastic mechanics, Hertz model described the relationship between the normal contact force $F_N$ and the indentation depth $\delta$, as shown in Eq. (1),

$$F_N = k\delta^n \qquad (1)$$

where $k$ represents the generalized stiffness parameter and $\delta$ is the indentation caused by deformation, as illustrated in Fig.1b. The exponent $n$ depends on the topological properties of the contacting surfaces [28]. Theoretical analysis shows that for the contacts between sphere, cube, prism, cylinder(horizontal and vertical ), cone and elastic half space, $n$ are 1.5, 1.0, 1.0, 1.0, 1.0 and 2.0 respectively [27], as shown in Fig.2. The spherical surface is the most representative among them [7], so in most studies $n$ is considered to be 1.5. Of course, this is not applicable in all situations, for instance, $n$ is considered approximately equal to 2.0 for grains with irregular surfaces like sand in some cases [29]. Generally speaking, the exponent *n* is variable for different kinds of contacting spaces.

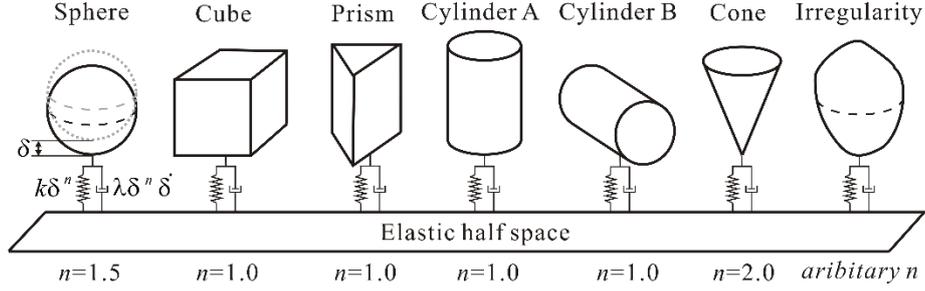

Fig. 2 Equivalent system to the contact problems with multiple contacting surfaces.

Although the energy dissipation during collision is neglected by Hertz model which is derived from the foundation of the elasticity theory, the literature [30] pointed out that even though 40% of the kinetic energy is lost due to viscous dissipation during the collision, Hertz's theory is still accurate in predicting the collision time and the maximum contact area. Therefore the Hertz model can be used as a basis for the contact problems with energy dissipation, and several continuous contact models considering energy dissipation during collision have been proposed. In these models all energy losses that occur during the collision are assumed to be taken by the viscous damper [7, 31] and the contact system is equivalent to a single degree of freedom dynamic system with nonlinear spring and nonlinear damping, as shown in Fig. 2. The contact force is composed of elastic force and dissipative force, and the exponent of damping coefficient is generally considered to be equal to the exponent of the indentation depth which is proposed by Hunt and Crossley [21], consequently the expression of contact force is,

$$F_N = k\delta^n + \lambda\delta^n\dot{\delta} \tag{2}$$

where $\lambda$ is the hysteresis damping factor, theoretical analysis shows that $\lambda$ is a function of the equivalent stiffness $k$, the coefficient of restitution, and the initial relative velocity. Compared to the Kelvin-Voigt model with linear damping, the models with nonlinear damping are more accurate and avoid nonzero contact force at zero deformation. So the expression form of the contact force as described in Eq. (2) is considered as the basis of most continuous contact force models [1, 6, 20], also in our study.

## 3. The energy loss associated with the restitution coefficient

For more accurate modeling, energy change during the collision process need to be considered. The description of the energy change becomes facilitated by the utilizing of the coefficient of restitution which is easy to measure [7], although it can not describe the details of the collision process, including contact force, deformation, et al.

The coefficient of restitution, denoted $c_r$, is defined as

$$c_r = -\frac{\dot{\delta}^{(+)}}{\dot{\delta}^{(-)}} \tag{3}$$

where $\dot{\delta}^{(-)} = v_i^{(-)} - v_j^{(-)}$ and $\dot{\delta}^{(+)} = v_i^{(+)} - v_j^{(+)}$ are the initial relative velocity and the relative separating velocity respectively.

According to the energy balance and the law of conservation of linear momentum in the period from $t^{(-)}$ to $t^{(+)}$, the following expressions are acquired

$$\left[\frac{1}{2}m_i\left(v_i^{(-)}\right)^2 + \frac{1}{2}m_j\left(v_j^{(-)}\right)^2\right] = \Delta E_{loss} + \left[\frac{1}{2}m_i\left(v_i^{(+)}\right)^2 + \frac{1}{2}m_j\left(v_j^{(+)}\right)^2\right] \tag{4}$$

$$m_i v_i^{(-)} + m_j v_j^{(-)} = m_i v_i^{(+)} + m_j v_j^{(+)} \tag{5}$$

The expression of the total amount of energy loss $\Delta E_{loss}$ can be obtained through Eqs. (4) and (5):

$$\Delta E_{loss} = \frac{1}{2}m\left[\left(v_i^{(-)} - v_j^{(-)}\right)^2 + \left(v_i^{(+)} - v_j^{(+)}\right)^2\right] \tag{6}$$

where $m$ is the equivalent mass of the contact system which defined as $m = \frac{m_i m_j}{m_i + m_j}$.

Combined with the definition of the restitution coefficient $c_r$ as described in Eq. (3), Eq. (6) can be simplified as

$$\Delta E_{loss} = \frac{1}{2}m(1 - c_r^2)\dot{\delta}^{(-)^2} \tag{7}$$

**4. The construction of the new model**

4.1. General issues of the construction of the new model

Although the coefficient of restitution can describe the changes of energy and velocity before and after the collision concisely, but some important details of the contact-impact process can not be obtained as mentioned above. Previous studies have shown that continuous contact models can depict these details as the hysteresis damping factor $\lambda$ is fixed by the measured value of restitution coefficient. As one continuous contact model was adopted to simulate the contact-impact process, the simulation value of restitution coefficient can be calculated by the simulation values of velocities before and after the collision. In this research, the set value of restitution coefficient which is provided as input to the contact force models for calculating the hysteresis damping factor is defined as pre-restitution coefficient, and the simulation value of restitution coefficient is defined as post-restitution coefficient. From the view of physical point, the smaller the difference between the post and pre-restitution coefficients, the closer the collision process described by the model to the actual situation [9], and consequently more accurate description of the dynamic behavior of multibody systems.

According to the expression of impact force described in Eq. (2), the system dynamic equation is represented below,

$$m\ddot{\delta} + k\delta^n + \lambda\delta^n\dot{\delta} = 0 \tag{8}$$

where gravity is ignored as it is much smaller than the impact force.

The determination of the expression of $\lambda$ is the key step of the construction of the contact model. It is almost impossible to obtain analytical solution to Eq. (8), several ways have been applied to derive the approximate expression of damping coefficient $\lambda$ in previous studies [1, 7, 9, 17, 21-25]. In recent continuous contact model studies, literature [7] and [9] provided new contact force models respectively in the light of the approximate functions between deformation velocity and indentation depth for the case of n=1.5, the models are much closer to the actual situation than other models in terms of the difference between post and pre-restitution coefficients [1, 9], especially for soft materials with low restitution coefficients [7].

Inspired by the above research of constructing contact force model, this article try to give a new continuous contact model on the foundation of an approximate dynamic equation for contact problems with regular or irregular contacting surfaces, this means that the exponent $n$ is arbitrary.

4.2. Approximate dynamic equation for the impact system

The system dynamic equation as expressed in Eq. (8) can be rewritten as

$$m\ddot{\delta} + (k + \lambda\dot{\delta})\delta^n = 0 \tag{9}$$

Let us consider an equation similar to Eq. (9):

$$m\ddot{\delta} + (k + \lambda\hat{\dot{\delta}})\delta^n = 0 \tag{10}$$

where $\hat{\dot{\delta}}$ is the equivalent velocity, which keeps a constant during compression or restitution phases and associated with $\dot{\delta}^{(-)}$, the details will be discussed in the following section.

By integrating Eq. (10) over the compression phase, it can be derived that,

$$\dot{\delta}^2 - \dot{\delta}_0^2 = -\frac{2(k + \lambda\hat{\dot{\delta}})}{m(n+1)}\delta^{n+1} \tag{11}$$

If $\dot{\delta} = 0$, $\delta = \delta_m$, Substituting $\dot{\delta} = 0$ into Eq.(11) we have

$$\delta_m^{n+1} = \frac{m(n+1)}{2(k + \lambda\hat{\dot{\delta}}_c)}\dot{\delta}^{(-)^2} \tag{12}$$

The function between deformation velocity $\dot{\delta}$ and indentation depth $\delta$ during compression phase can be acquired from Eq. (12),

$$\dot{\delta} = \dot{\delta}^{(-)}\sqrt{1 - \left(\frac{\delta}{\delta_m}\right)^{n+1}} \tag{13}$$

Then applied it to calculate the work done by dissipative force during compression phase according to Eq. (2):

$$\Delta E_c = \int_0^{\delta_m} \lambda\delta^n\dot{\delta}d\delta = \int_0^{\delta_m} \lambda\delta^n\dot{\delta}^{(-)}\sqrt{1 - \left(\frac{\delta}{\delta_m}\right)^{n+1}}d\delta$$

$$= \int_0^{\delta_m} \frac{\lambda\delta_m^{n+1}\dot{\delta}^{(-)}}{n+1}\sqrt{1 - \left(\frac{\delta}{\delta_m}\right)^{n+1}}\,d\left(\frac{\delta}{\delta_m}\right)^{n+1} \tag{14}$$

$$= \int_0^1 \frac{\lambda\delta_m^{n+1}\dot{\delta}^{(-)}}{n+1}\sqrt{1-x}\,dx = \frac{\lambda\delta_m^{n+1}}{n+1} \cdot \frac{2\dot{\delta}^{(-)}}{3}$$

On the other hand, as for the similar dynamic equation as illustrated in Eq. (10), the work done by dissipative force during compression phase is,

$$\Delta E_c^* = \int_0^{\delta_m} \lambda\delta^n\hat{\dot{\delta}}_c d\delta = \frac{\lambda\delta_m^{n+1}}{n+1}\hat{\dot{\delta}}_c \tag{15}$$

Combining Eqs. (14) and (15), the equivalent velocity during compression phase can be deduced:

$$\hat{\dot{\delta}}_c = \frac{2\dot{\delta}^{(-)}}{3} \tag{16}$$

Similarly with the derivation of compression phase, by integrating Eq. (10) over the restitution phase, we obtain

$$\left(c_r\dot{\delta}^{(-)}\right)^2 - \dot{\delta}^2 = \frac{2(k + \lambda\hat{\dot{\delta}}_r)}{m(n+1)}\delta^{n+1} \tag{17}$$

Substituting $\dot{\delta} = 0$ into Eq. (17) it can be deduced that,

$$\delta_m^{n+1} = \frac{m(n+1)}{2(k + \lambda\hat{\dot{\delta}}_r)}\left(c_r\dot{\delta}^{(-)}\right)^2 \tag{18}$$

Noting that the velocities have opposite direction during restitution phase and compression phase, the function between deformation velocity $\dot{\delta}$ and indentation depth $\delta$ during restitution phase can be derived:

$$\dot{\delta} = -c_r\dot{\delta}^{(-)}\sqrt{1 - \left(\frac{\delta}{\delta_m}\right)^{n+1}} \tag{19}$$

The work done by dissipative force during restitution phase is,

$$\Delta E_r = \int_{\delta_m}^0 \lambda\delta^n\dot{\delta}d\delta = -\int_{\delta_m}^0 \lambda\delta^n c_r\dot{\delta}^{(-)}\sqrt{1 - \left(\frac{\delta}{\delta_m}\right)^{n+1}}d\delta = \frac{\lambda\delta_m^{n+1}}{n+1} \cdot \frac{2c_r\dot{\delta}^{(-)}}{3} \tag{20}$$

As for the similar dynamic equation described in Eq. (10), the work done by dissipative force during restitution phase can be gained,

$$\Delta E_r^* = \int_{\delta_m}^{0} \lambda \delta^n \hat{\dot{\delta}}_r d\delta = -\frac{\lambda \delta_m^{n+1}}{n+1} \hat{\dot{\delta}}_r \qquad (21)$$

Combining Eqs. (20) and (21), the equivalent velocity of restitution phase can be expressed as

$$\hat{\dot{\delta}}_r = \frac{-2c_r \dot{\delta}^{(-)}}{3} \qquad (22)$$

In summary, the approximate dynamic equation can be described as follows:

$$\begin{cases} m\ddot{\delta} + \left(k + \lambda \frac{2c_r \dot{\delta}^{(-)}}{3}\right)\delta^n = 0 & \text{Compression phase} \\ m\ddot{\delta} + \left(k - \lambda \frac{2c_r \dot{\delta}^{(-)}}{3}\right)\delta^n = 0 & \text{Restitution phase} \end{cases} \qquad (23)$$

The above discussion suggests that the approximate dynamic equation with properly chosen velocities can well approximate the system dynamic equation. There will be more discussion of this aspect in following sections.

4.3. The expression of hysteresis damping factor

Based on the approximate dynamic equation, the expression of contact force can be proposed for contact-impact problems with irregular contacting surfaces and the exponent $n$ for the system dynamic equation is arbitrary.

Combining with Eqs. (14) and (20), we have

$$\frac{\Delta E_c}{\Delta E_r} = \frac{1}{c_r} \qquad (24)$$

Due to $\Delta E = \Delta E_c + \Delta E_r$, substituting Eq. (24) into Eq. (7) yields

$$\Delta E_c = \frac{m\dot{\delta}^{(-)^2}(1-c_r)}{2} \qquad (25)$$

According to the energy balance and the law of conservation of linear momentum in the period from $t^{(-)}$ to $t_m$, we can get following equations:

$$\begin{cases} \frac{m_i(v_i^-)^2}{2} + \frac{m_j(v_j^-)^2}{2} = \frac{(m_i + m_j)v_{ij}^2}{2} + \Delta E_k + \Delta E_c \\ m_i v_i^- + m_j v_j^- = (m_i + m_j)v_{ij} \end{cases} \qquad (26)$$

$\Delta E_k$ is the elastic potential energy stored up from $t^{(-)}$ to $t_m$, it can be calculated as

$$\Delta E_k = \int_{0}^{\delta_m} k\delta^n d\delta = \frac{k\delta_m^{n+1}}{n+1} \qquad (27)$$

Combining Eqs. (25), (26) and (27), by denoting the equivalent mass $m = \frac{m_i m_j}{m_i + m_j}$ and $\dot{\delta}^{(-)} = v_i^- - v_j^-$, it can be deduced that,

$$\frac{m\dot{\delta}^{(-)^2}}{2} = \frac{k\delta_m^{n+1}\dot{\delta}^{(-)^2}}{n+1} + \frac{m\dot{\delta}^{(-)^2}(1-c_r)}{2} \qquad (28)$$

Then yields

$$\delta_m^{n+1} = \frac{c_r m(n+1)\dot{\delta}^{(-)^2}}{2k} \qquad (29)$$

Substituting Eq. (29) into Eqs. (12) and (18), we get

$$\Delta E_c = \frac{\lambda c_r m \dot{\delta}^{(-)^3}}{3k} \qquad (30)$$

$$\Delta E_r = \frac{\lambda c_r^2 m \dot{\delta}^{(-)^3}}{3k} \qquad (31)$$

Combining Eqs. (7), (30) and (31), we can deduced that,

$$\frac{\lambda c_r m \dot{\delta}^{(-)3}}{3k} + \frac{\lambda c_r^2 m \dot{\delta}^{(-)3}}{3k} = \frac{m(1 - c_r^2)\dot{\delta}^{(-)2}}{2} \tag{32}$$

Then the description of hysteresis damping factor can be obtained,

$$\lambda = \frac{3k(1-c_r)}{2\dot{\delta}^{(-)} - 3(1-c_r)\frac{2\dot{\delta}^{(-)}}{3}} = \frac{3k(1-c_r)}{2c_r \dot{\delta}^{(-)}} \tag{33}$$

We can notice that this description is as the same as the model provided in [9] which is just considered n=1.5 for the system dynamic equation.

Substituting $\hat{\dot{\delta}}_c = \frac{2\dot{\delta}^{(-)}}{3}$, $\hat{\dot{\delta}}_r = \frac{-2c_r\dot{\delta}^{(-)}}{3}$, and Eq.(33) into Eqs. (12) and (18) which are derived from the approximate dynamic equation, respectively, we obtain

$$\delta_m^{n+1} = \frac{m(n+1)\left(\dot{\delta}^{(-)}\right)^2}{2\left(k + \frac{3k(1-c_r)}{2c_r\dot{\delta}^{(-)}} \cdot \frac{2\dot{\delta}^{(-)}}{3}\right)} = \frac{c_r m(n+1)\dot{\delta}^{(-)2}}{2k} \tag{34}$$

$$\delta_m^{n+1} = \frac{m(n+1)\left(c_r\dot{\delta}^{(-)}\right)^2}{2\left(k - \frac{3k(1-c_r)}{2c_r\dot{\delta}^{(-)}} \cdot \frac{2c_r\dot{\delta}^{(-)}}{3}\right)} = \frac{c_r m(n+1)\dot{\delta}^{(-)2}}{2k} \tag{35}$$

It can be seen that Eqs. (34) and (35) are completely consistent with Eq. (29) which is derived according to the energy balance and the law of conservation of linear momentum. And the feasibility of establishing new model based on the approximate dynamic equation is further confirmed.

4.4. Establishing the new model by modifying

Although the derivations in section 4.2 and 4.3 demonstrate that the approximate dynamic equation achieves a good approximation of the system dynamic equation in terms of the system energy consumption and the maximum indentation depth, however, it is worth discussing the approximation degree of the approximate equation to the system dynamic equation in the contact-impact process.

The damping coefficient is calculated by using Eq. (33) and then substituted it into Eqs. (2) and (23), a series of numerical experiments is conducted to evaluate the approximation degree of the approximate dynamic equation to the dynamic behavior of the system, as shown in Figs. 3 and 4. It can be seen that the approximate equation is not perfect for the approximation of the system dynamics equation. There is a deviation in the description of the system dynamic behavior between the approximate dynamic equation and the system dynamic equation, and the smaller the coefficient of restitution $c_r$, the larger the deviation. Naturally, we can speculate that it is possible to obtain a more accurate model by modifying the formula as shown in Eq. (33) which is developed based on the approximate equation.

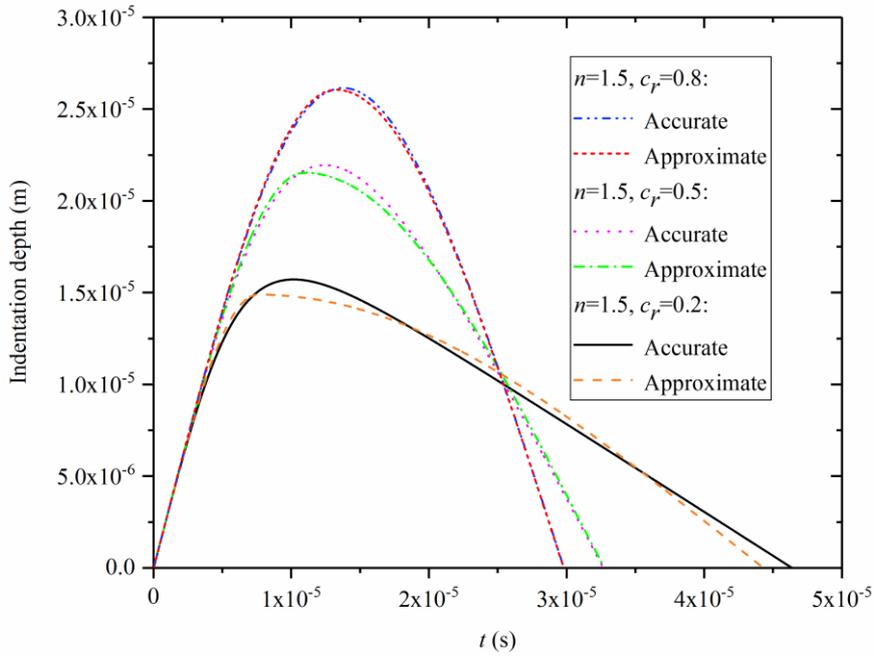

Fig.3 Dynamic behaviors of the system dynamic equation and the approximate dynamic equation as $n$=1.5.

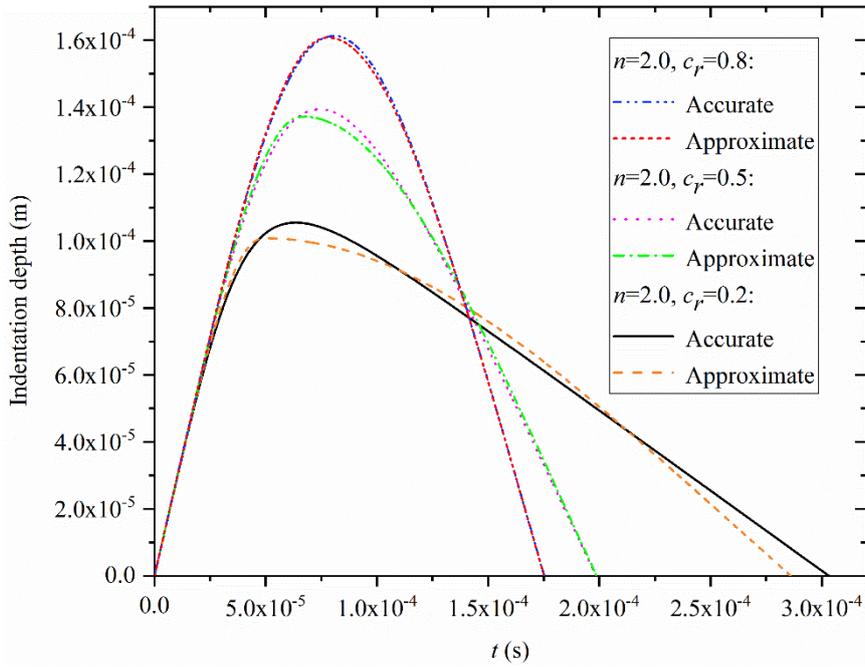

Fig.4 Dynamic behaviors of the system dynamic equation and the approximate dynamic equation as $n$=2.0.

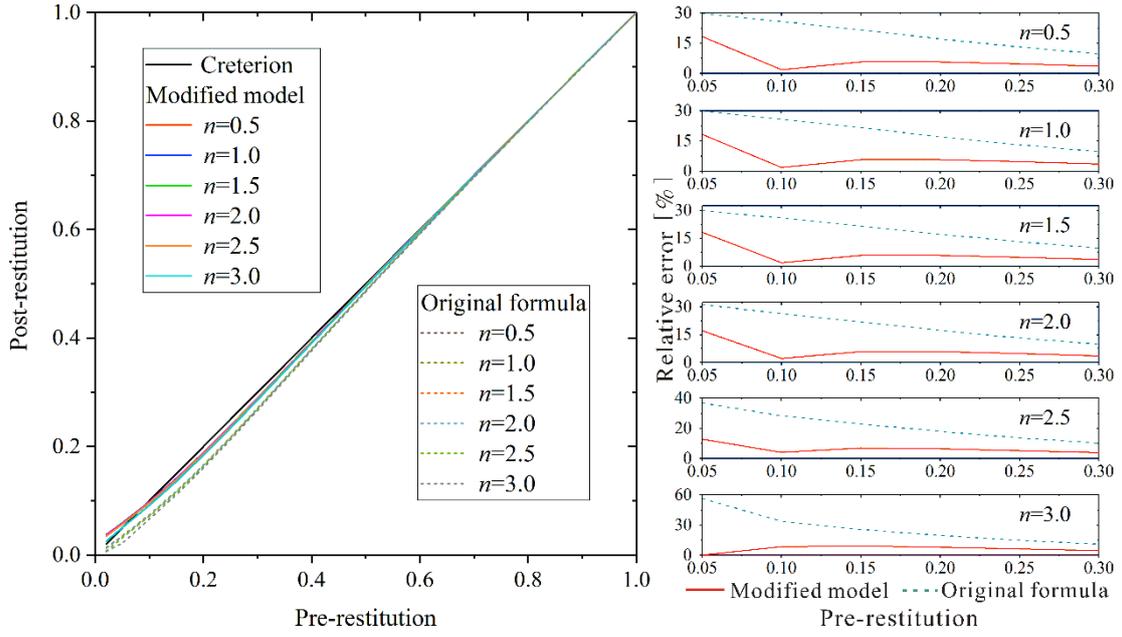

Fig.5 Relation and relative errors between the post and pre-restitution coefficients for modified model and original formula.

For continuous contact force models, previous studies have shown that the pre-restitution coefficient differs from the post-restitution coefficient, which should theoretically be the same [1, 7, 9]. As the pre-restitution coefficient is obtained via physical experiment, the deviation of the post-restitution coefficient from the pre-restitution coefficient means the deviation of the simulation results of velocity from actual values of velocity, and the deviation of the output predicted by the model from actual behavior of the system. So the deviation of the post-restitution coefficient from the pre-restitution coefficient is considered as an important criteria for the modifying of the model. Through a large number of numerical simulations, a more accurate contact model with less deviation of post-restitution coefficient from pre-restitution coefficient is established by modifying the original formula of the hysteresis damping factor describing in Eq. (33), through the minor adjustment of the equivalent velocity and trial calculations. The plots of post vs. pre-restitution coefficient for modified model and original formula are presented in Fig.5. The hysteresis damping factor in the new model is given as follows:

$$\lambda = \frac{3k(1-c_r)}{2\dot{\delta}^{(-)} - 3(1-c_r)\frac{160\dot{\delta}^{(-)}}{249}} = \frac{249k(1-c_r)}{6\dot{\delta}^{(-)} + 160c_r\dot{\delta}^{(-)}} \tag{36}$$

It can be noted that for the purely elastic contact, the hysteresis damping factor $\lambda$ equals to zero. Correspondingly, the contact force in the new model can be expressed as

$$F_N = k\delta^n + \frac{249k(1-c_r)}{6\dot{\delta}^{(-)} + 160c_r\dot{\delta}^{(-)}}\delta^n\dot{\delta} \tag{37}$$

The new contact-impact model described in Eqs. (36) and (37) is valid for direct central and frictionless impacts with regular or irregular contacting surfaces as the exponent *n* in the system dynamic equation which characterizes the topological properties of the contacting surfaces is arbitrary. The partial relative errors between post and pre-restitution coefficient for modified model and original formula are also given in Fig.5, it shows that, relative to the original formula, the modified model can improve the accuracy effectively.

Through a serious of numerical simulations, the plots of post vs. pre-restitution coefficient for different contact force models as $n=1.5$ are given in Fig. 6. It can be seen that the new model provided the best fit for the post vs. pre-restitution coefficient relative to other models as $n=1.5$, especially for the low value of restitution coefficient. Combined with the results of Fig. 5, it can be inferred that this conclusion is still valid for other values of $n$.

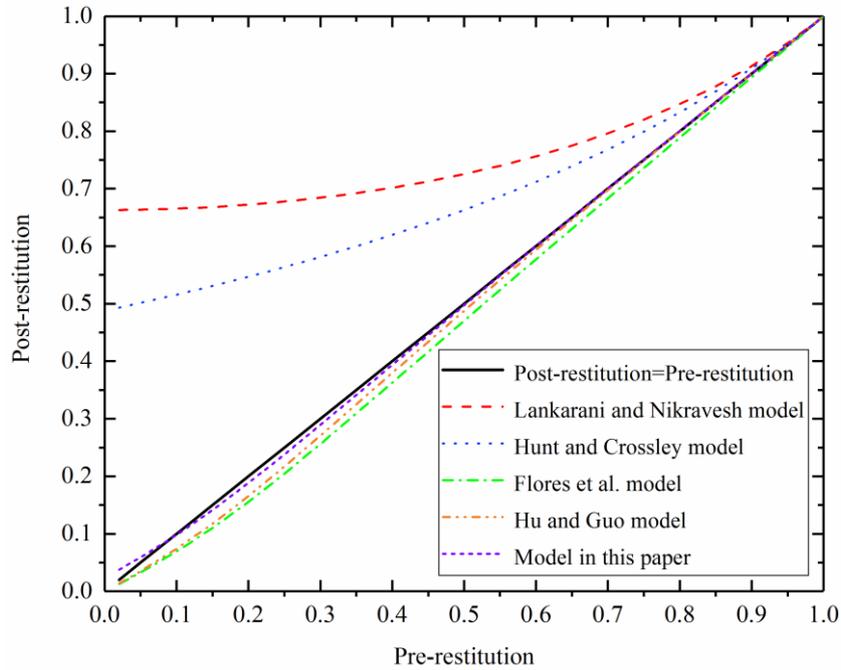

Fig.6 Relation between the post and pre-restitution coefficients for different contact models.

## 5. Numerical simulations

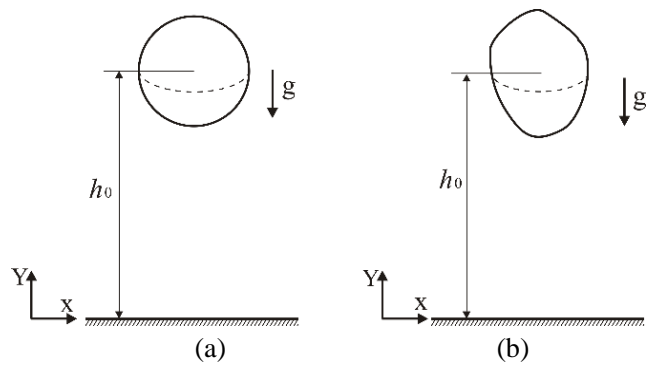

Fig.7 Impact of falling object (a) an elastic sphere, (b) an elastic object with irregular contacting surface.

To evaluate the validity of the contact model proposed in this study and evaluate the influences of the geometries of contacting surfaces on the system dynamic responses, the contact-impact of a free-falling object is simulated. As shown in Fig. 7, an elastic body made of PTFE release from the initial position which under the influence of gravity only, and collides with the ground which is assumed to be rigid and stationary. The basic parameters of the simulation are given in Table 1 [7, 9]. Several numerical simulations are conducted by utilizing Matlab codes when the coefficient of restitution is set as 0.9 and 0.2 respectively.

Firstly, an elastic sphere with radius 0.1m is considered and the exponent $n$ is set as 1.5 correspondingly, as illustrated in Fig. 7a. Five different continuous contact models are utilized for modeling the first contact-impact event of the free-falling sphere, including the models developed by Lankarani and Nikravesh [17], Hunt and Crossley [21], Flores et al. [7], Hu and Guo

[9], and the model described in this work. The time histories of the deformation and contact force of the ball during the contact-impact process are demonstrated in Figs. 8-11. In addition, Table 2 provides the values of post and pre-restitution coefficients of five different contact force models. It can be seen that when the restitution coefficient is 0.9, the simulation results on the foundation of the five models are basically the same, and the differences between the post-restitution coefficients calculated by using each model and the pre-restitution coefficients are also very small.

Table 1
The basic parameters of the simulations.

| Characteristics | Values |
| --- | --- |
| Mass of the falling object | $m = 1.0$ kg |
| Initial height | $h_0 = 1.0$ m |
| Gravity acceleration | $g = 9.8$ m/s$^2$ |
| Equivalent stiffness of the contact | $K = 1.4 \times 10^8$ N/m$^{3/2}$ |
| Initial velocity of the falling object | $v_0 = 0$ m/s |

As illustrated in Figs. 10 and 11, when the restitution coefficient is 0.2, the simulation results calculated by Flores et al. model, Hu and Guo model and the model described in this paper tend to be consistent with little differences. In contrast, calculations from the Lankarani and Nikravesh model and the Hunt and Crossley model lead to greater indentation depth, smaller contact force, and shorter contact duration, with a relative deviation of approximately 25%-45% from the results of the three models mentioned above. As demonstrated in Table 2, it can be seen that the post-restitution coefficients calculated based on Flores et al. model, Hu and Guo model and the model described in this paper are relatively close to the pre-restitution coefficients, while the post-restitution coefficients obtained by Lankarani and Nikravesh model and Hunt and Crossley model are quite different from the pre-restitution coefficients. It can be inferred that Flores et al. model, Hu and Guo model and the model described in this paper are not only suitable for the case of low energy dissipation during contact process, but also suitable for the case of higher energy dissipation. In another word, these three models can perform well for the entire range of the coefficient of restitution.

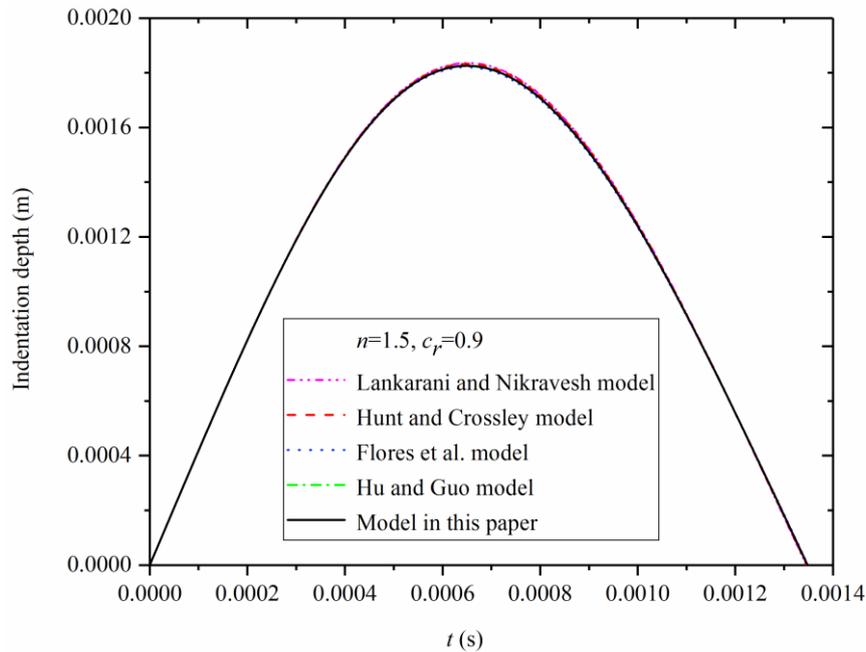

Fig.8 Time history of the indentation depth for different contact force models as $c_r$=0.9.

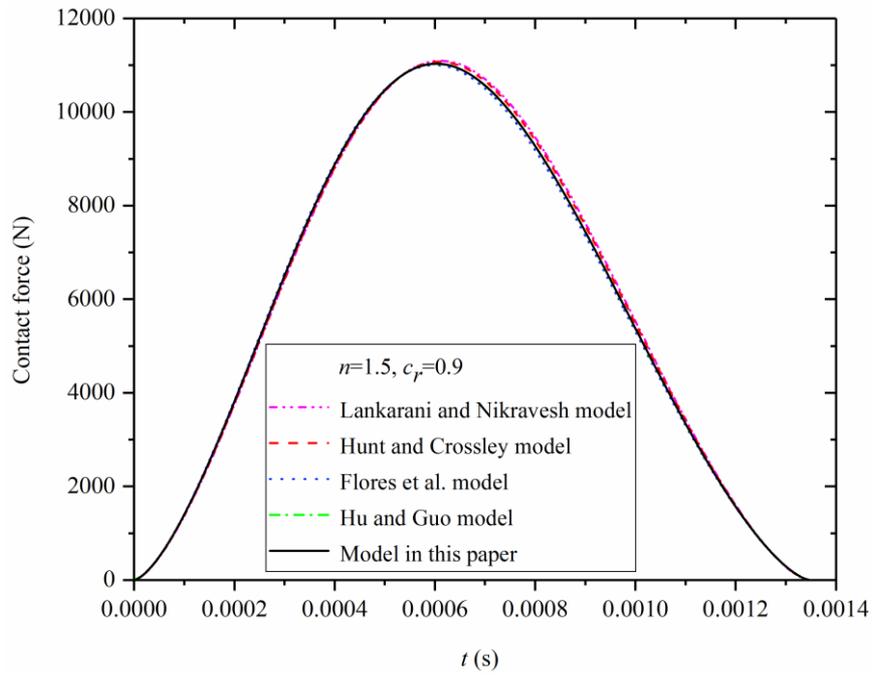

Fig.9 Time history of the contact force for different contact force models as $c_r$=0.9.

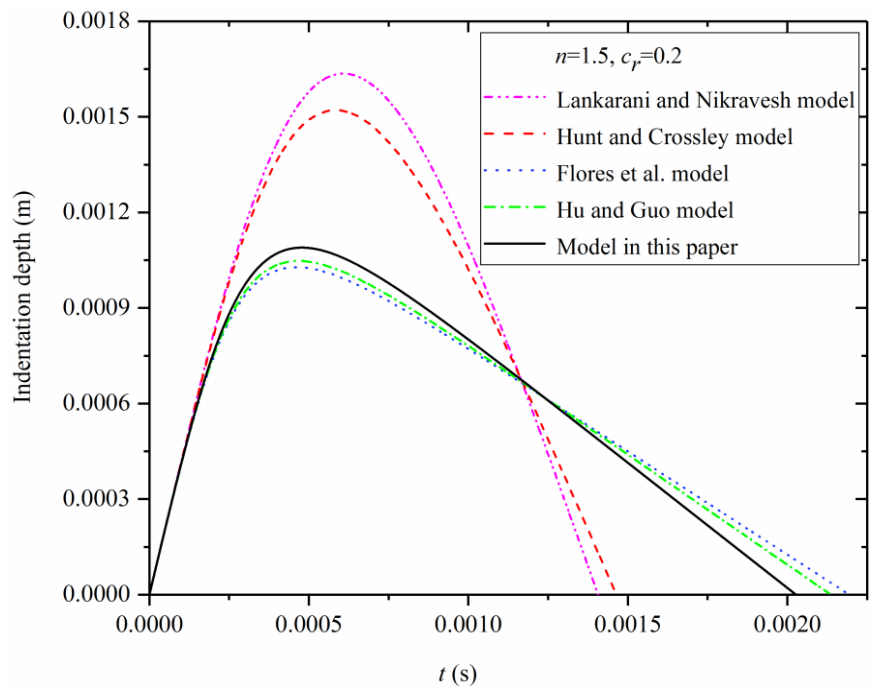

Fig.10 Time history of the indentation depth for different contact force models as $c_r$=0.2.

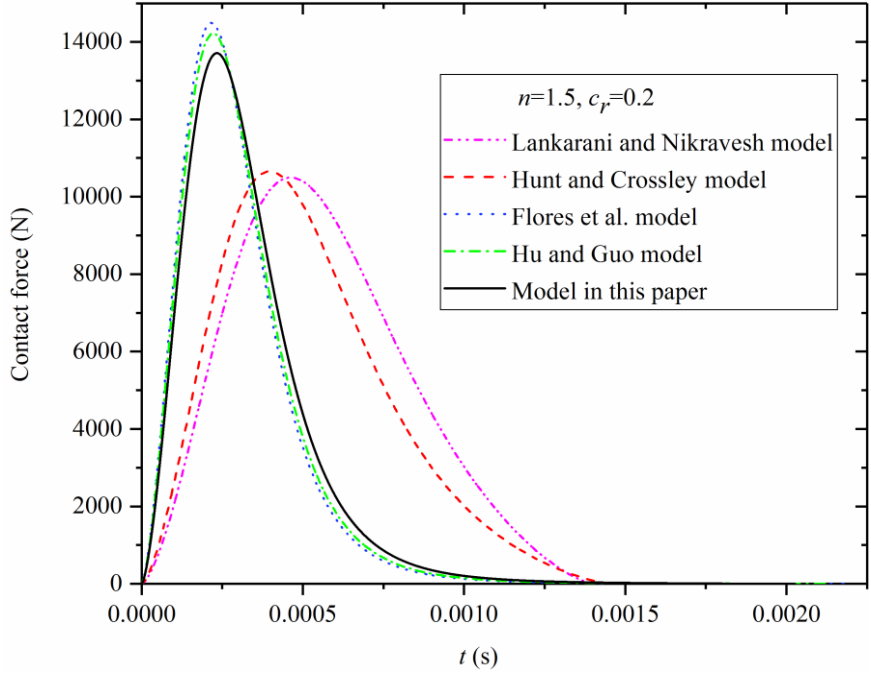

Fig.11 Time history of the contact force for different contact force models as $c_r$=0.2.

Table 2
Post and pre-restitution coefficients for contact force models.

| pre-restitution coefficient | post-restitution coefficient | Lankarani and Nikravesh model [17] | Hunt and Crossley model [21] | Flores et al. model [7] | Hu and Guo model [9] | The model in this paper |
|---|---|---|---|---|---|---|
| 0.9 | | 0.9129 | 0.9088 | 0.8937 | 0.8997 | 0.9000 |
| Relative error[%] | | 1.44% | 0.98% | 0.70% | 0.04% | 0.001% |
| 0.2 | | 0.6715 | 0.5459 | 0.1537 | 0.1638 | 0.1867 |
| Relative error[%] | | 235.75% | 172.92% | 23.12% | 18.09% | 6.65% |

Then the model described in this paper is applied to simulate the contact-impact of a free-falling object with irregular contacting surface. The numerical simulations are conducted with the initial condition as shown in Table 1 and the exponent $n$ is chosen as 1.0, 1.2 and 1.5 respectively which describes the topological properties of the contacting surfaces.

As illustrated in Figs. 12-15, it can be noticed that under the same other conditions, the simulation results of the collision process are obviously different under small differences of values of $n$. With the rise of $n$, the indentation depth and contact-impact time increase, and the maximum contact force decreases. Moreover, it can be seen that with the reduction of the restitution coefficient, the symmetry between the compression phase and the restitution phase becomes weaker. As shown in Figs. 14 and 15, when $c_r$=0.2, the duration of the restitution phase is significantly longer than that of the compression phase, and the differences between the occurrence time of the maximum contact force and the maximum indentation depth during the contact-impact process are more obvious. It also supports the significance of developing the contact force model for contact problems with regular or irregular contacting surfaces.

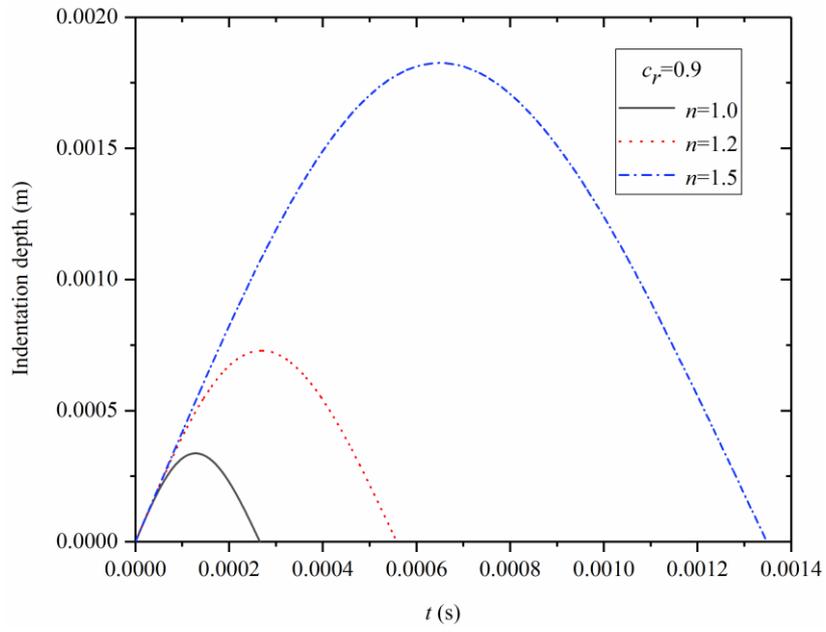

Fig.12 Time history of the indentation depth for different contacting surfaces as $c_r$=0.9.

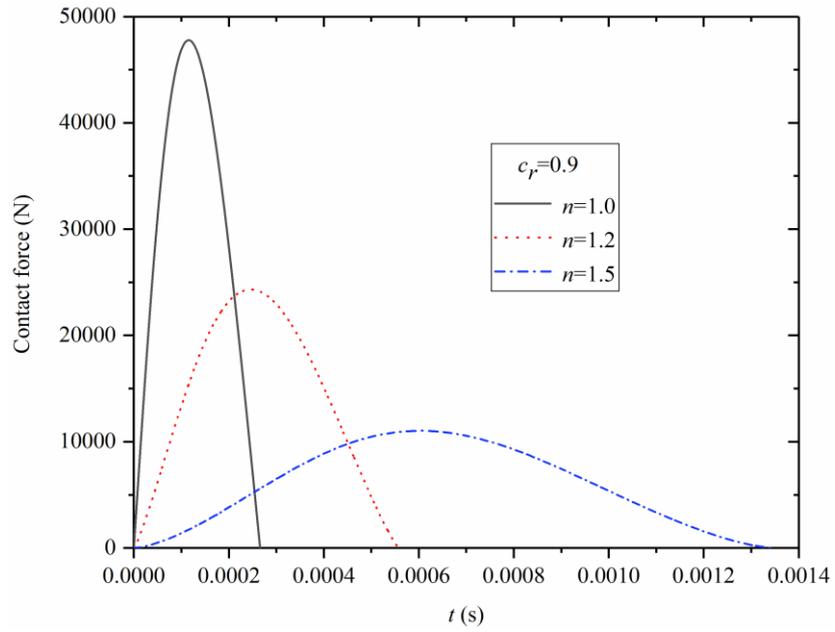

Fig.13 Time history of the contact force for different contacting surfaces as $c_r$=0.9.

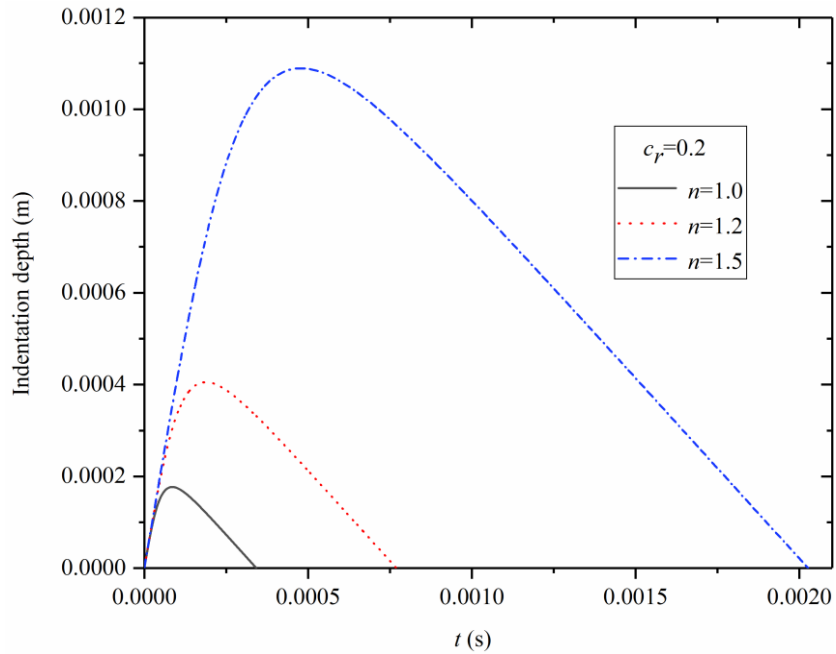

Fig.14 Time history of the indentation depth for different contacting surfaces as $c_r$=0.2.

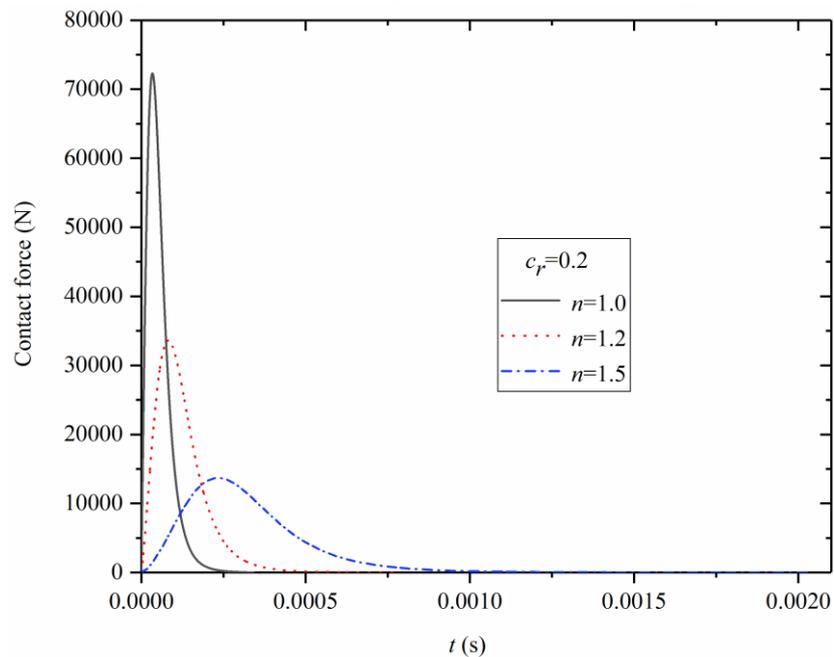

Fig.15 Time history of the contact force for different contacting surfaces as $c_r$=0.2.

## 6. Conclusions

A new continuous contact force model has been put forward for impact analysis in multibody dynamics with regular or irregular contacting surfaces and with energy dissipations. The model is proposed on the foundation of the Hertz law and a hysteresis damping force is introduced for modeling the energy dissipation. The main difficulty of this research is that it is almost impossible to obtain an analytical solution from the system dynamic equation. An approximate dynamic equation is developed by introducing equivalent velocity. The analysis and simulations illustrate that the approximate dynamic equation with properly chosen equivalent velocities can well approximate the system dynamic equation. An approximate function between deformation velocity and deformation is obtained in the light of the approximate dynamic equation, and it is utilized to calculate the energy loss due to the damping force. Then the expression of the hysteresis damping parameter is derived by combining the energy balance and the law of conservation of linear momentum. Considering the deviation between the approximate equation and the system dynamic equation, a more accurate model is established by modifying the original formula of the hysteresis damping parameter.

Numerical results of five different continuous contact models reveal the capability of the new model. By comparing the results of the new model, the Flores et al. model and Hu and Guo model, we know that these three have quite similar responses for the low and high values of the restitution coefficient, which means they can perform well for soft and hard contact problems. Furthermore, the new model provided the best fit for the post vs. pre-restitution coefficient relative to other models, especially for the low value of restitution coefficient. The new model provides us a valuable method to simulate the contacting problems of various shapes, numerical results reveal that the influences of the geometry of the contacting surfaces on the dynamic system responses are significant, which demonstrates that the presentation of the new model is of great value. Like the Flores et al. model and Hu and Guo model, the new model can be utilized for impact analysis of a multibody mechanical system conveniently.

**Acknowledgements** This research was supported by the National Key Research and Development Plan of China under Grant No. 2019YFB1300200, the National Natural Science Foundation of China under Grant No. 11702294 and Beijing Natural Science Foundation under Grant No. 3194047.